\documentclass{iau}
\usepackage{graphicx}
\usepackage{natbib}
\usepackage{color}

\title[Protoplanetary Disks: Singles vs. Binaries] %% give here short title %%
{Protoplanetary Disk Evolution:\\Singles vs.\ Binaries}

\author[S. Daemgen et al.]   %% give here short author list %%
{Sebastian Daemgen$^1$,
Ray Jayawardhana$^2$,
Monika G. Petr-Gotzens$^3$,
 \and Elliot Meyer$^1$
}

\affiliation{$^1$Department of Astronomy \& Astrophysics, University of Toronto, 50 St. George Street, Toronto, ON, Canada M5H 3H4, email: {\tt daemgen@astro.utoronto.ca}\\
$^2$Faculty of Science, 4700 Keele Street, Toronto, ON M3J 1P3, Canada\\
$^3$European Southern Observatory, Karl-Schwarzschildstr.\ 2, 85748, Garching, Germany}

\pubyear{2015}
\volume{314}  %% insert here IAU Symposium No.
\pagerange{119--126}
\jname{Young Stars \& Planets Near the Sun}
\editors{J. H. Kastner, B. Stelzer, \& S. A. Metchev, eds.}
\begin{document}

\maketitle

\begin{abstract}
Based on a large number of observations carried out in the last decade it appears that the fraction of stars with protoplanetary disks declines steadily between $\sim$1\,Myr and $\sim$10\,Myr. We do, however, know that the multiplicity fraction of star-forming regions can be as high as $>$50\% and that multiples have reduced disk lifetimes on average. As a consequence, the observed roughly exponential disk decay can be fully attributed neither to single nor binary stars and its functional form may need revision. Observational evidence for a non-exponential decay has been provided by \citet{kra12}, who statistically correct previous disk frequency measurements for the presence of binaries and find agreement with models that feature a constantly high disk fraction up to $\sim$3\,Myr, followed by a rapid ($\lesssim$2\,Myr) decline. 

We present results from our high angular resolution observational program to study the fraction of protoplanetary disks of single and binary stars separately. We find that disk evolution timescales of stars bound in close binaries ($<$100\,AU) are significantly reduced compared to wider binaries. The frequencies of accretors among single stars and wide binaries appear indistinguishable, and are found to be lower than predicted from planet forming disk models governed by viscous evolution and photoevaporation.
\keywords{Stars: pre-main sequence; Stars: formation; circumstellar matter; binaries: general}
\end{abstract}

\firstsection % if your document starts with a section,
\section{Introduction}
The formation of gas giant planets requires significant amounts of gas and dust to be present in the circumstellar environment of a young T\,Tauri star.
The lifetime of protoplanetary disks is accordingly an important observable to constrain planet formation. 
To infer disk lifetimes, a number of previous studies have targeted young star-forming regions to measure the fraction of stars that exhibit either ongoing accretion or hot circumstellar dust or both. These fractions appear to be a strong function of the age of a star-forming region, monotonically decreasing from $\gtrsim$80\% to 0\% within $\sim$10\,Myr \citep[e.g.,][]{jay06,fed10}. The functional shape appears to be fit well by an exponential decay with a time constant of $\tau\approx2$--3\,Myr \citep{fed10}.

The fact that a large fraction of all young stars is bound in multiple systems \citep{duc13} has a strong effect on the conclusions that can be drawn from this observation because a) the presence of stellar binary companions has a strong effect on disk evolution, and b) a large number of binary stars typically remains undetected and will ``contaminate'' single star measurements.

\noindent\paragraph{\bf The effect of stellar binarity on disk evolution.}
Sub-mm observations show that multiple stars exhibit a much reduced total dust mass compared to undisturbed systems \citep{har12,and13}. This can be explained by the dynamical interaction of a disk with the stellar companion. For wide binaries this leads to a truncation of the outer disk to $\sim$1/5--1/2 times the binary separation \citep{art94}. This implies shorter disk lifetimes, given that mass accretion rates in binaries are comparable to those of single stars \citep{whi01,dae12}. 
And indeed, the data of \citet[see also \citealt{kra12}]{dae12,dae13} imply an exponential decay of binary circumstellar disk fractions with a time constant of $\tau\approx0.9$--1.3\,Myr, much shorter than that found by previous studies that do not focus on binary stars. 

The fact that disks in binary stars follow a different evolutionary path may help to constrain planet formation scenarios: \citet{duc10} found that stars with close ($\rho<100$\,AU) binary companions are less likely to be orbited by a low-mass ($\lesssim$1\,M$_\mathrm{Jup}$) gas planet than wider binaries and single stars. One possible interpretation calls for a slow gas planet-forming process for planets $<$1\,M$_\mathrm{Jup}$ -- too slow to complete in close binaries before the disk disperses. 
More massive planets, in contrast, may form through a rapid process that completes equally often around single stars and around even closely separated binary components where disk lifetimes are shorter.

\noindent\paragraph{\bf The difficulty of identifying single stars.}
While above considerations and 
the fact that sample sizes are typically small 
%small sample sizes 
paint a noisy but consistent picture of disk evolution in T\,Tauri \emph{binaries}, the evolution of a disk in an undisturbed system like that of a single star cannot be easily observed. This is partly due to our lack of knowledge about the true multiplicity status of young stars: 
efficient high-angular resolution imaging searches for companions do not reach much below $\sim$0.1$^{\prime\prime}$ ($\sim$10\,AU at the distance of the nearest star-forming regions of $\sim$100\,pc), and few young stars have been subject to radial velocity (RV) monitoring over baselines $>$10\,yr (equivalent to $\gtrsim$5\,AU). 
This leaves the 5--10 AU separation range mostly unstudied for stellar companions in the best-sampled nearby regions, and typically the range of unassessed separations is even larger. 

Owing to this observational effect, the fraction of undetected binary companions in any previous disk frequency survey can be $\sim$30\% or larger, depending on the distance of the targeted region and the availability of suitable survey data \citep[\emph{submitted}]{kra12,dae15}. The previously derived disk fractions in these samples accordingly represent a mixture of binary and single star disk data compromising their potential to provide physical information about disk decay.

\noindent\paragraph{\bf Disk evolution around single stars.}
Applying a statistical correction for undetected binarity, \citet{kra12} study the early evolution of single stars and find observational evidence for a disk fraction evolution that may be different from an exponential decay. In qualitative agreement with disk evolution models by \citet{ale09}, disk frequencies stay close to 100\% until $\sim$2--3\,Myr of stellar evolution and then rapidly decay before $\sim$6\,Myr of age. 
Such a change would have strong implications for planet formation. For example, gas giant planets may form on much longer timescales -- or start to form at a later time -- than typically assumed. 

While providing interesting qualitative evidence, the results reported by \citet{kra12} are based on unresolved binaries with no information about whether circumstellar material is distributed around one or both stars. Furthermore, their study relies on the sum of a variety of disk indicators which probe different physical aspects of a disk. The values presented by \citet{kra12} accordingly represent upper limits to the true accretion and inner dust disk frequency of single stars.

To gain a better understanding of the evolution of undisturbed disk material in the first few Myr of stellar evolution, a consistent and direct measurement of disk frequency around single stars is needed. As current instruments do not enable the identification of binary companions at all separations and mass ratios, the frequency of undetected binary companions and the disk fraction in multiples must be taken into careful consideration to infer single star values. In the following we describe our previous and ongoing efforts.

\section{A coherent analysis of disks in single and multiple stars}
We measured the frequency of disks around 52 spatially resolved multiple stars with separations between $\sim$25--1000\,AU in the Orion Nebula Cluster and Chamaeleon\,I star-forming regions \citep{dae12,dae13}. For each stellar component, our measurements trace both the presence of ongoing accretion based on Brackett-$\gamma$ (Br$\gamma$) emission as well as hot inner dust inferred from near-infrared color excess.

We compare this binary data set to single stars in Chamaeleon\,I \citep[\emph{submitted}]{dae15}. In order to compile a sample with as little binary contamination as possible, we queried the high-angular resolution imaging survey by \citet{laf08} for stars without stellar companions between $\sim$50--500\,AU and brighter than $\Delta K_\mathrm{s}\approx3$\,mag. Stars with known radial velocity companions were removed from the sample. 
A total of 54 single star candidates were observed with SOFI/NTT $K$-band spectroscopy to assess the presence of emission at the wavelength of Br$\gamma$. Using a Monte Carlo simulation, we simulated the overall companion distribution of our sample and find that, even after excluding all adaptive optics (AO) and RV companions, there remains a $\sim$30\% chance of multiplicity for any star in the sample. Using the results from our Chamaeleon\,I binary study \citep{dae13}, we correct the measured single star accretion frequency for the bias introduced by undetected binarity.

As these measurements are based on a well-defined set of suitable disk indicators (Br$\gamma$, NIR color excess) and good coverage by AO and RV companion searches, the resulting disk evolution measurements represent the most robust assessment of the relative abundance of disks around the individual components of binaries and single stars to date.

\section{Results}
\begin{figure}[t]
\begin{center}
\includegraphics[width=0.78\textwidth]{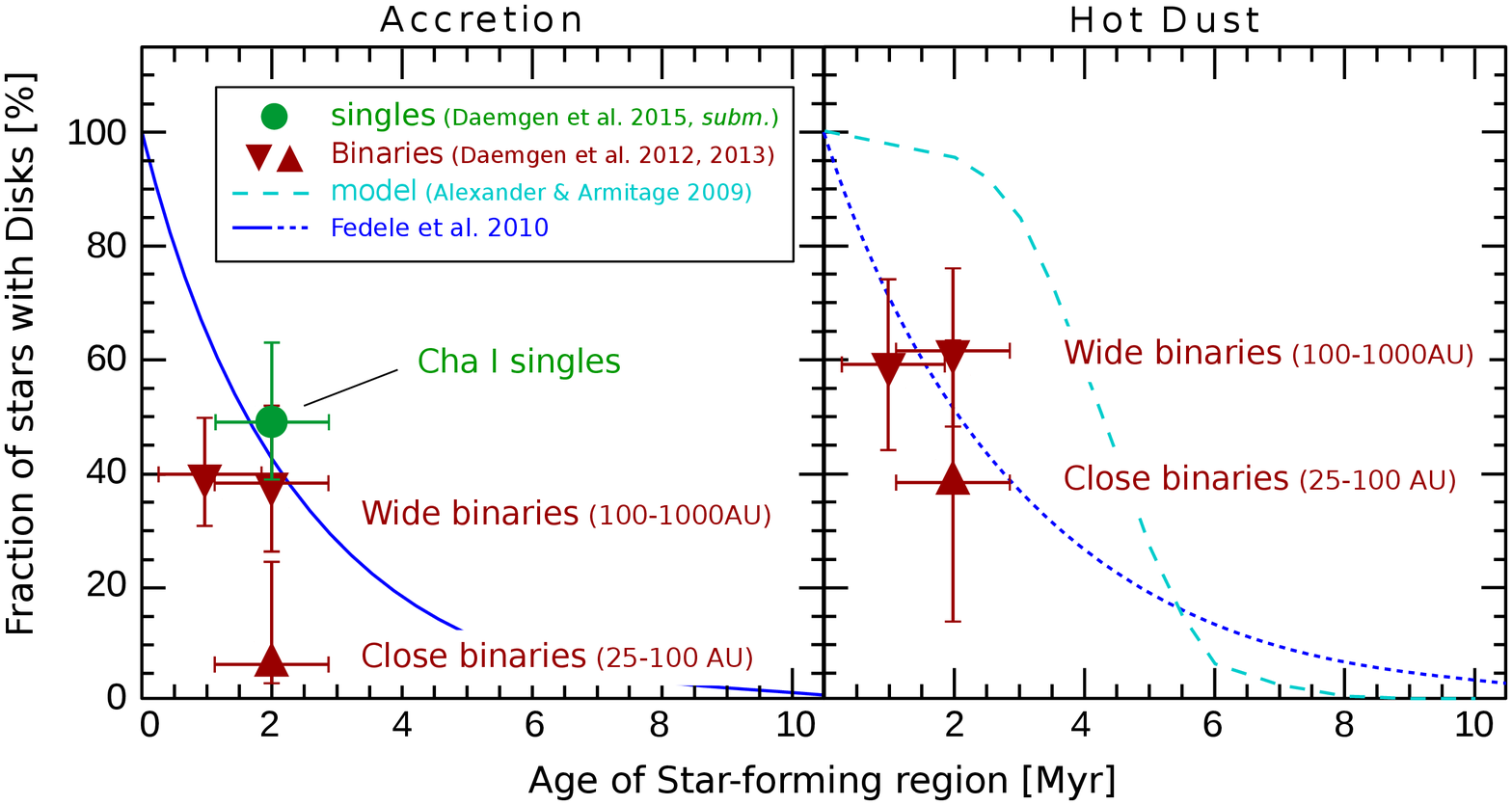}
 \caption{Disk frequency of single stars and components of binary stars in the Orion Nebula Cluster (1\,Myr) and Chamaeleon\,I (2\,Myr). \emph{Left:} accretion inferred from Brackett-$\gamma$ emission. \emph{Right:} hot circumstellar dust as inferred from $H$--$K$ and $K$--$L^\prime$ excess \citep{dae12,dae13,dae15}. The blue continuous and dotted lines show exponential decay functions measured by \citet[$\tau_\mathrm{accr}\sim2.3$\,Myr, $\tau_\mathrm{dust}\sim3$\,Myr]{fed10}, not corrected for undetected multiplicity. The dashed cyan line shows the disk evolution model by \citet{ale09}.}
   \label{fig1}
\end{center}
\end{figure}

Figure~\ref{fig1} shows our assessment of single and binary star disk fraction as a function of age. Close binaries ($\rho\le100$\,AU) appear to disperse their disks on much smaller timescales than wider binaries, both for our accretion and hot inner disk measurements.

We furthermore robustly confirm that the inferred single star accretor fraction \citep[$F=48^{+14}_{-10}$\%;][\emph{submitted}]{dae15} is about 6 times larger than that of close binaries. The single star accretor fraction appears to be slightly larger but consistent with both wide binaries and the decay curve found by previous surveys without binary correction \citep[e.g.,][]{fed10}. In particular, our new measurement of the single star accretor fraction is less than 1$\sigma$ larger than previous estimates of the accretion frequency in Chamaeleon\,I \citep[$\sim$41\%, based on][]{dam07}. 
The data disfavor a scenario where all stars retain their disks until 2--3\,Myr \citep{ale09}. 

It has recently been suggested that ages of star forming regions are much older than previously assumed \citep{bel13,som15}. If the ages of the targeted Orion Nebula Cluster and Chamaeleon\,I regions turn out to be a factor of $\sim$2 larger than assumed here, then agreement between the data and the model is preserved.

\section{Summary and Conclusions}
Single stars are hard or even impossible to identify in most star-forming regions. 
This is because contamination with binary companions that are inaccessible to today's observational methods is on the order of $\gtrsim$30\% for a typical star-forming region like Chamaeleon\,I.
As close binaries ($<$100\,AU) have been shown to exhibit disk evolution timescales that are shorter than those of wide binaries and single stars, typical disk frequency studies are biased by undetected binary companions. 
Rather than following an exponential decay, as suggested by studies without binary correction, the fraction of single stars with disks may stay constant and close to 100\% until $\sim$2--3\,Myr, followed by a rapid ($\lesssim$3\,Myr) drop, according to disk evolution models that include viscosity, photoevaporation, and planet formation.

We conducted the first dedicated single star disk frequency measurement in Cha\-maeleon\,I. We find a significantly higher accretor fraction among single stars than around components of close binary stars ($<$100\,AU) in the same region. 
The single star disk frequency appears to be inconsistent with the very high disk frequency at 2\,Myr suggested by previous models and statistical considerations. Part of the discrepancy between the observations and models might be due to systematic age uncertainties.

More measurements of single and binary star disk frequencies are needed to infer the evolution of undisturbed disks as a function of time and to constrain the qualitative and quantitative evolution of disk material in both truncated and undisturbed disks.

\end{document}